\documentclass[11pt]{article}
\usepackage[margin=0.8in]{geometry}
\usepackage{amsmath,amssymb,cases}
\usepackage{siunitx}
\usepackage{booktabs}
\usepackage{hyperref}
\usepackage{caption}

\usepackage{graphicx}

\title{Temporal metamaterials with passive switching as impedance-matched absorbers}
\author{Suat Bar{\i}\c{s} \.Iplik\c{c}io\u{g}lu\\
	Department of Electrical and Electronics Engineering, Koç University\\
	İstanbul, Türkiye\\
	\texttt{siplikcioglu15@ku.edu.tr}}
\date{}

\begin{document}
	\maketitle
	
\begin{abstract}
	Recent experiments on temporal reflection in transmission line metamaterials and theoretical treatments of dispersive time-varying media have established the fundamental role of modulation mechanisms in the interface conditions, underpinning the introduction of passive photonic time crystals with stable momentum band gaps. Drawing from these concepts, it is shown that temporal metamaterials with simultaneous passive permittivity and permeability switching exhibit wideband absorption with impedance-matching, effectively behaving as one-dimensional perfectly matched layers. Under the effective medium theory, the loss mechanism is attributed to the emergent effective electric and magnetic conductivities, which are used to derive an approximate matching condition for asynchronous modulation and to engineer lossy material properties. The proposed approach and its performance beyond the Rozanov bound are verified with semi-analytical calculations as well as full-wave simulations, and the limits on realizing a two-dimensional temporal perfectly matched layer are theoretically evaluated.
\end{abstract}
\maketitle

\section{Introduction}
Dynamic control over temporal interfaces in complex media has unlocked a new regime of electromagnetic phenomena~\cite{Mostafa2024}, ranging from parametric amplification~\cite{Asgari2024} and inverse prisms~\cite{Akbarzadeh2018} to temporal diffraction~\cite{Tirole2023} and spatiotemporal wave control~\cite{PachecoPena2020a,PachecoPena2020b}. To date, these interfaces have been explored across a diverse array of media, revealing complex rich physical behavior in plasmas~\cite{Felsen1970,Stepanov1976,Kalluri2005,Li2022}, Lorentzian media~\cite{Stanic1991,Solis2021,Bakunov2021a,Bakunov2021b,Bakunov2024,Dixon2026}, chiral materials~\cite{Yin2022,Mostafa2023}, layered or wire metamaterials~\cite{Ptitcyn2025,Sidorenko2025,Simovski2025} and non-Foster structures~\cite{PachecoPena2025}. Beyond theoretical studies, temporal interfaces were experimentally realized with dynamic transmission lines (TLs) in microwaves~\cite{Moussa2023,Jones2024,Kuusela2025,Hou2026} and with epsilon-near-zero materials under a pump-probe configuration in optics~\cite{Tirole2023,Zhou2020,Lustig2023}. The impact of this paradigm extends beyond electromagnetics, with analogous temporal phenomena recently being observed in acoustic~\cite{Kim2024} and water waves~\cite{Apffel2022}.

The development of these time-varying systems has prompted a re-examination of the boundary conditions that characterize wave behavior at a temporal interface~\cite{Xiao2014,Bakunov2014,Galiffi2025}. Recent experiments on time-modulated TLs with passive switching of the distributed capacitance~\cite{Moussa2023} have highlighted the validity of different boundary conditions during on- and off-switching~\cite{Galiffi2025,Maslov2025}. Within the context of unsteady plasmas and Lorentzian media, similar distinct boundary conditions were also  derived to characterize wave dynamics during rapid ionization and charge recombination~\cite{Stepanov1976,Bakunov2021a,Bakunov2021b,Galiffi2025,Bernety2026}. Concurrent with these interpretations of the temporal interfaces, the concept of passive photonic time crystals (PTC) with stable momentum band gaps was introduced for dynamic TL metamaterials~\cite{Yin2024} and Lorentzian media~\cite{Bakunov2024}.

This work demonstrates that a temporal metamaterial with impedance-matched passive permittivity and permeability switching, analogous to a dynamic TL medium, exhibits linear dispersion and constant loss profile without any momentum band gaps, showcasing properties similar to those of a one-dimensional perfectly matched layer (PML)~\cite{Berenger1994}. Relying on high-frequency temporal modulations in tandem with passive and reactive switching, the matched absorption mechanism presented herein is distinct from the previous time-variant configurations, which are based on backward wave cancellation~\cite{PachecoPena2020b}, tapered or stepped temporal transitions~\cite{Pan2025,Hou2026}, spectral redistribution/trapping of wave energy through single-shot switching~\cite{Shlivinski2018,Li2020,Firestein2022,Yang2022},
coherent wave control~\cite{Suwunnarat2019,Galiffi2023,Galiffi2026} or resistive/active Floquet modulations~\cite{Chambers2005,Mostafa2022,Hayran2024,Ciabattoni2025}. Furthermore, since the absorption is largely facilitated by high modulation frequencies, it does not require a phase synchronization with the incident waveform. An effective medium analysis~\cite{Pachecopena2019,Torrent2020} of such materials is also shown to reveal effective electric and magnetic conductivities that can be used to engineer lossy material properties, as well as relaxed matching conditions for asynchronous modulation. The limitations on extending the approach to the realization of a multidimensional PML are also theoretically evaluated. The results are verified with semi-analytical calculations and full-wave simulations.

\section{Theory}
\subsection{Temporal interfaces in switched media}
The main starting point of the analyses herein is the TL metamaterial medium. As in the conventional microwave theory~\cite{Kinayman2005}, TL metamaterials can be modeled as simple LC ladders but with time-dependent distributed capacitances and inductances, which is equivalent to an isotropic medium with a capacitance-dependent permittivity and inductance-dependent permeability~\cite{Yin2023}. Capacitance modulation is achieved by opening or closing an ideal switch between two parallel capacitors, whereas inductance modulation is performed through opening or closing a switch that shorts one of the two series inductors. The case of capacitance or inductance increase can be treated by conventional flux continuity boundary conditions, as reported by Morgenthaler in 1958~\cite{Morgenthaler1958}; nevertheless, the reverse case requires a first-principles treatment. When the parallel capacitor is switched off, bound charges are lost from the system, decreasing energy and momentum; such temporal interfaces can be represented with the continuity of voltage, rather than charge~\cite{Galiffi2025}. Importantly, the corresponding switch should be connected to the ground node to discharge the capacitor after switching. The dual problem is the shorting of a series inductor with a closed switch: this case corresponds to the loss of magnetic flux linkage and can be represented by continuity of current, rather than magnetic flux~\cite{Maslov2025}.
\begin{figure}[h!]
	\centering 
	\includegraphics[width=0.75\textwidth]{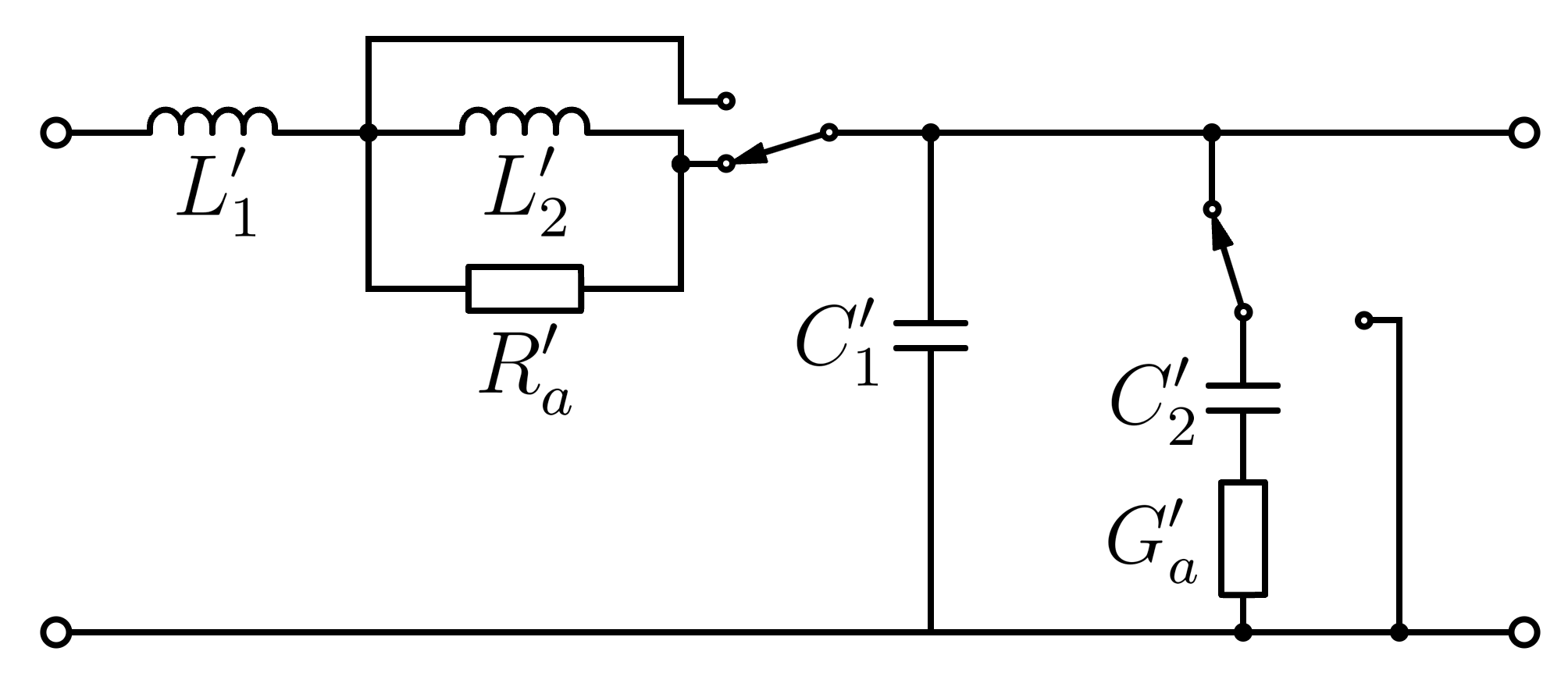}
	\caption{Distributed element representation of a transmission line under reactance switching with series inductors and parallel capacitors. Single-pole double-throw switches on $L_2'$ and $C_2'$ prevent the reintroduction of residual charges and fluxes during periodic modulation, whereas the auxiliary elements $R_a'$ and $G_a'$ act as damping paths while balancing the wave impedance for all frequencies. The primed quantities represent per-unit-length densities.} \label{fig1}
\end{figure}

While this description accurately describes the single-shot switching events, periodic modulation of the reactances requires a more intricate treatment of the equivalent circuit. Shorting of the inductor or disconnecting the capacitor does not eliminate the flux linkage or charges but merely makes them unavailable to the propagating wave: the former results in the excitation of a DC current that circulates in the closed loop~\cite{Maslov2025}, whereas the bound charges remain in the capacitor. This requires the modification of the dynamic transmission line topology using the distributed element model shown in Fig~\ref{fig1}, with primes denoting the quantities per unit length. Under this model, a single-pole double-throw switching configuration with auxiliary resistive elements $R_a'$ and $G_a'$ is employed to safely dissipate the remnant charges and fluxes in between multiple switching events, with the former decaying as $e^{-t R_a' / L_2'}$ and the latter as $e^{-t G_a' / C_2'}$. While the charges on $C_2'$ can be removed in principle without $G_a'$ by simply grounding it, $R_a'$ alone introduces an artificial mismatch and frequency dispersion in the form of Debye relaxation~\cite{Scarlatti2001}; thus $G_a'$ restores the balance. In the high-reactance state, requiring the characteristic impedance to be frequency-independent yields the condition:
\begin{align}
	\frac{L_1' + L_2'}{C_1' + C_2'} = \frac{L_1' + \frac{L_2'}{1 - i\omega \frac{L_2'}{R_a'}}}{C_1' + \frac{C_2'}{1 - i \omega \frac{C_2'}{G_a'}}}
\end{align}
It should be noted that this condition is only satisfied by impedance-matched transitions under $\tau_a=L_2'/R_a' = C_2'/G_a'$, where $\tau_a$ is the shared time constant of switched reactances. This is analogous to the Heaviside condition in TLs~\cite{Kinayman2005}. While this perfectly matches the waves across the temporal interface for all frequencies, it does not alleviate the Debye dispersion of the propagating waves in between the transitions. The ensuing dispersion can be managed with the appropriate choice of a large $R_a'$ and $G_a'$, ensuring time constants much smaller than both the duration between switching events and the period of the propagating wave: well-posed nondispersive and passive temporal interface conditions are then retrieved for $\omega\tau_a \ll 1$. In turn, the parasitic loss tangent due to the auxiliary resistances is $O(\omega\tau_a)$ and vanishes at small time constants, being distinct from energy-removing passive transitions. Another major point to note is that each resistive element provides a physical path for energy decrease during the on-switching, resolving the two-capacitor and two-inductor paradoxes~\cite{Maslov2025}.

Alternative to the given configuration, a switched TL can be constructed with an LC ladder of two parallel inductors and two series capacitors, in which the energy decrease is attributed to the excitation of static magnetic and electric modes in lieu of flux or charge removal~\cite{Maslov2025}, necessitating similar treatment for periodic modulation.

Proceeding henceforth with analogous macroscopic electrodynamic quantities strictly in the nondispersive limit, the loss of charges and fluxes at a temporal interface with decreasing reactances at $t=t_0$ can be represented by an impulse of electric $(\textbf{J} \propto \delta(t-t_0))$ and magnetic currents $(\textbf{K} \propto \delta(t-t_0))$~\cite{Galiffi2025} in Maxwell's curl equations:
\begin{align}
	\int_{t_0^-}^{t_0^+} dt \; \left( \nabla \times \textbf{E} \right) &= 
	- \int_{t_0^-}^{t_0^+} dt \left( \frac{d\textbf{B}}{dt} + \textbf{K}\right) = 0 \\
	\int_{t_0^-}^{t_0^+} dt \; \left( \nabla \times \textbf{H} \right) &=  \int_{t_0^-}^{t_0^+} dt \left( \frac{d\textbf{D}}{dt} + \textbf{J}\right) = 0 
\end{align}
The current terms can be alternatively decomposed as:
\begin{align}
	\textbf{B}(t=t_0^+)-\textbf{B}(t=t_0^-)&= 
	-\mu_0\int_{t_0^-}^{t_0^+} dt \Delta \mu \delta(t-t_0) \underbrace{\frac{1}{2} \left[\textbf{H}(t=t_0^+)+\textbf{H}(t=t_0^-)\right]}_{\textbf{H}(t=t_0)} \\
	\textbf{D}(t=t_0^+)-\textbf{D}(t=t_0^-) &= -\varepsilon_0\int_{t_0^-}^{t_0^+} dt \Delta \varepsilon \delta(t-t_0)  \underbrace{\frac{1}{2} \left[\textbf{E}(t=t_0^+)+\textbf{E}(t=t_0^-)\right]}_{\textbf{E}(t=t_0)} 
\end{align}
where $\Delta \varepsilon$ and $\Delta \mu$ are the absolute differences in relative permittivities and permeabilities, respectively. Thus, the impulsive electric and magnetic conductivities are in the form $\sigma(t)=\varepsilon_0 \Delta \varepsilon \delta(t-t_0)$ and $\sigma^*(t)=\mu_0 \Delta \mu \delta(t-t_0)$. For smooth temporal transitions, these can be represented as~\cite{Galiffi2025}:
\begin{numcases}{\sigma(t) =}
	-\varepsilon_0 \frac{d\varepsilon(t)}{dt}, & $\frac{d\varepsilon(t)}{dt}<0$ \label{eq:smooth-1e} \nonumber
	\\
	0, & $\frac{d\varepsilon(t)}{dt} \geq 0$ \label{eq:smooth-2e}
\end{numcases}
\begin{numcases}{\sigma^*(t) =}
	-\mu_0 \frac{d\mu(t)}{dt}, & $\frac{d\mu(t)}{dt}<0$ \label{eq:smooth-1m} \nonumber
	\\
	0, & $\frac{d\mu(t)}{dt} \geq 0$ \label{eq:smooth-2m}
\end{numcases}

\subsection{Impedance-matched photonic time crystals and effective medium theory}
It is known that under conventional temporal interface conditions~\cite{Morgenthaler1958}, a nondispersive and active PTC under synchronous permittivity and permeability modulation exhibits a linear dispersion without momentum band gaps as long as the wave impedance is preserved throughout the modulation~\cite{GaxiolaLuna2021,Dalarsson2025}. In comparison, a passive PTC with synchronous and impedance-matched modulation exhibits a net loss while having closed momentum band gaps, which can be represented with complex frequency eigensolutions in the form of $\omega=\omega_r+i\omega_i$ and $\omega_i < 0$ for all $k$ under the $e^{-i\omega t}$ time convention. The loss mechanism can be understood first by focusing on a single temporal unit cell of the time-periodic structure, e.g. a passive and matched temporal slab in a TL metamaterial. For this example, a simultaneous increase in both capacitance and inductance scales down the forward wave amplitude without causing any backward waves, while a coordinated decrease after a time delay ensures that the amplitude remains constant. In both cases, the instantaneous energy density is reduced, while no reflections occur. While the energy decrease at rising edges arises from the continuity of $\textbf{D}$ and $\textbf{B}$ that are in line with two-capacitor and two-inductor paradoxes~\cite{Galiffi2025,Maslov2025}, at falling edges, it stems from the disconnection of reactive elements, which instead enforces the continuity of $\textbf{E}$ and $\textbf{H}$. In electromagnetic terms, this can be interpreted by the effective material losses at the falling edges: when passive switching is employed, the propagating waves are introduced to an impulse train of electric and magnetic conductivities at the falling edges of the modulation, which compensate for the gain and introduce loss. Under the well-known matching condition for transverse electromagnetic (TEM) waves~\cite{Berenger1994}, no instantaneous impedance discontinuities and hence no reflections occur for synchronous modulation of permittivity and permeability at the falling modulation edges:
\begin{equation}
	\frac{\mu_0 \mu_r}{\varepsilon_0 \varepsilon_r} = \frac{\sigma^*}{\sigma} = \frac{\mu_0 \Delta \mu}{\varepsilon_0 \Delta \varepsilon} \label{eq:match}
\end{equation} 
Remarkably, this is equivalent to the Heaviside condition, with $\sigma$ and $\sigma^*$ being analogous to the shunt conductance and series resistance, respectively. This correspondence can be verified from the exact eigenproblem for the time-periodic and matched temporal slab: transfer matrix method (TMM) (Appendix \ref{appendixA}) reveals the closed-form dispersion relation to be:
\begin{align}
	\omega(k) = \pm k c_0 \frac{\varepsilon_1 \tau_2 + \varepsilon_2 \tau_1}{(\tau_1 + \tau_2) \varepsilon_1 \varepsilon_2} + \frac{2\pi m}{\tau_1 + \tau_2} + i \ln \left(\frac{\varepsilon_1}{\varepsilon_2}\right) \frac{1}{\tau_1 + \tau_2}, \;\; \text{for}\; \varepsilon_2 > \varepsilon_1 \; \text{and} \; m \in \mathbb{Z} \label{eq:ptc_matched_dispersion}
\end{align}
where $\varepsilon_{n}$ and $\tau_{n}$ are the relative permittivities and their durations. The dispersion relation herein is a linear function of $k$ while the loss is $k$-agnostic. It must be further stressed that temporal decay of the matched PTC is inversely proportional to the switching period, whereas the permittivity/permeability ratio affects it logarithmically.

Since synchronous and completely impedance-matched modulation is not always possible due to jitter effects, oscillator detuning and/or difficulties in obtaining a large permeability change, a more relaxed matching condition can be obtained through the lens of effective medium theory (EMT). The effective constitutive parameters for the aforementioned two-phase medium are given as~\cite{Pachecopena2019,Torrent2020}:
\begin{align}
	\varepsilon_{\text{eff}}&=\left[\frac{1}{T_e}\int_{0}^{T_e} dt \; \frac{1}{\varepsilon_r(t)} \right]^{-1}=\frac{T_e \varepsilon_1 \varepsilon_2}{\varepsilon_1 \tau_{2e} + \varepsilon_2 \tau_{1e}} \label{eq:emt-1}\\ 
	\mu_{\text{eff}}&=\left[\frac{1}{T_m}\int_{0}^{T_m} dt \; \frac{1}{\mu_r(t)} \right]^{-1}=\frac{T_m \mu_1 \mu_2}{\mu_1 \tau_{2m} + \mu_2 \tau_{1m}}  \label{eq:emt-2}
\end{align}
where $T_{e,m}=\tau_{1\{e,m\}}+\tau_{2\{e,m\}}$ is the modulation period of each constitutive parameter. Similarly, homogenization of the impulse train of electric and magnetic conductivities yields:
\begin{align}
	\sigma_{\text{eff}} & =\frac{1}{T_e}\int_{0}^{T_e} dt \; \sigma(t) = \frac{\varepsilon_0 \Delta \varepsilon }{T_e}  \label{eq:emt-3} \\
	\sigma_{\text{eff}}^* &=\frac{1}{T_m}\int_{0}^{T_m} dt \; \sigma^*(t) = \frac{\mu_0 \Delta \mu}{T_m}  \label{eq:emt-4}
\end{align}
Using Eqs.~\ref{eq:emt-1} to~\ref{eq:emt-4}, the approximate matching condition (per Eq.~\ref{eq:match}) under the long-wavelength limit can be represented as:
\begin{equation}
	\frac{\frac{\tau_{2e}}{\varepsilon_2}+\frac{\tau_{1e}}{\varepsilon_1}}{\frac{\tau_{2m}}{\mu_2}+\frac{\tau_{1m}}{\mu_1}}=\frac{T_e^2}{T_m^2}\frac{\Delta \mu}{\Delta \varepsilon} = \frac{T_e}{T_m} Z_c^2 \label{eq:emt_match}
\end{equation}
where $Z_c$ is the normalized wave impedance to be matched to. Besides matching, these effective medium parameters can be used to explain the behavior of passive PTCs with permittivity modulation, as well as to engineer effective dielectric or magnetic responses alongside loss in metastructures. This approach can also be interpreted as the effective medium generalization of the proposed formalism by Yin and Alù~\cite{Yin2024}, which designated duty cycle and impedance ratios as the characteristic variables for tuning momentum selectivity in passive PTCs, with the distributed capacitance being the tuning parameter. Significantly, the reciprocal of effective permittivity or equivalently permeability (Eqs.~\ref{eq:emt-1} and~\ref{eq:emt-2}) appears directly in the dispersion relation for matched PTC (Eq.~\ref{eq:ptc_matched_dispersion}), indicating that the fundamental mode index is governed by the effective constitutive parameters. This can be attributed to the fact that phase accumulates additively at temporal interfaces, whereas the amplitude scales multiplicatively. Thus, the homogenized conductivities in Eqs.~\ref{eq:emt-3} and~\ref{eq:emt-4} are essentially linearized with respect to the modulation depth, with the exact expression being:
\begin{align}
	\sigma_{\text{eff}} = \sigma_{\text{eff}}^* \frac{\varepsilon_0}{\mu_0} =  & -\varepsilon_0 \ln\left(\frac{\varepsilon_1}{\varepsilon_2}\right) \frac{\varepsilon_1 \varepsilon_2}{\varepsilon_1 \tau_2 + \varepsilon_2 \tau_1}
\end{align}

\begin{figure*}[t]
	\centering 
	\includegraphics[width=0.95\textwidth]{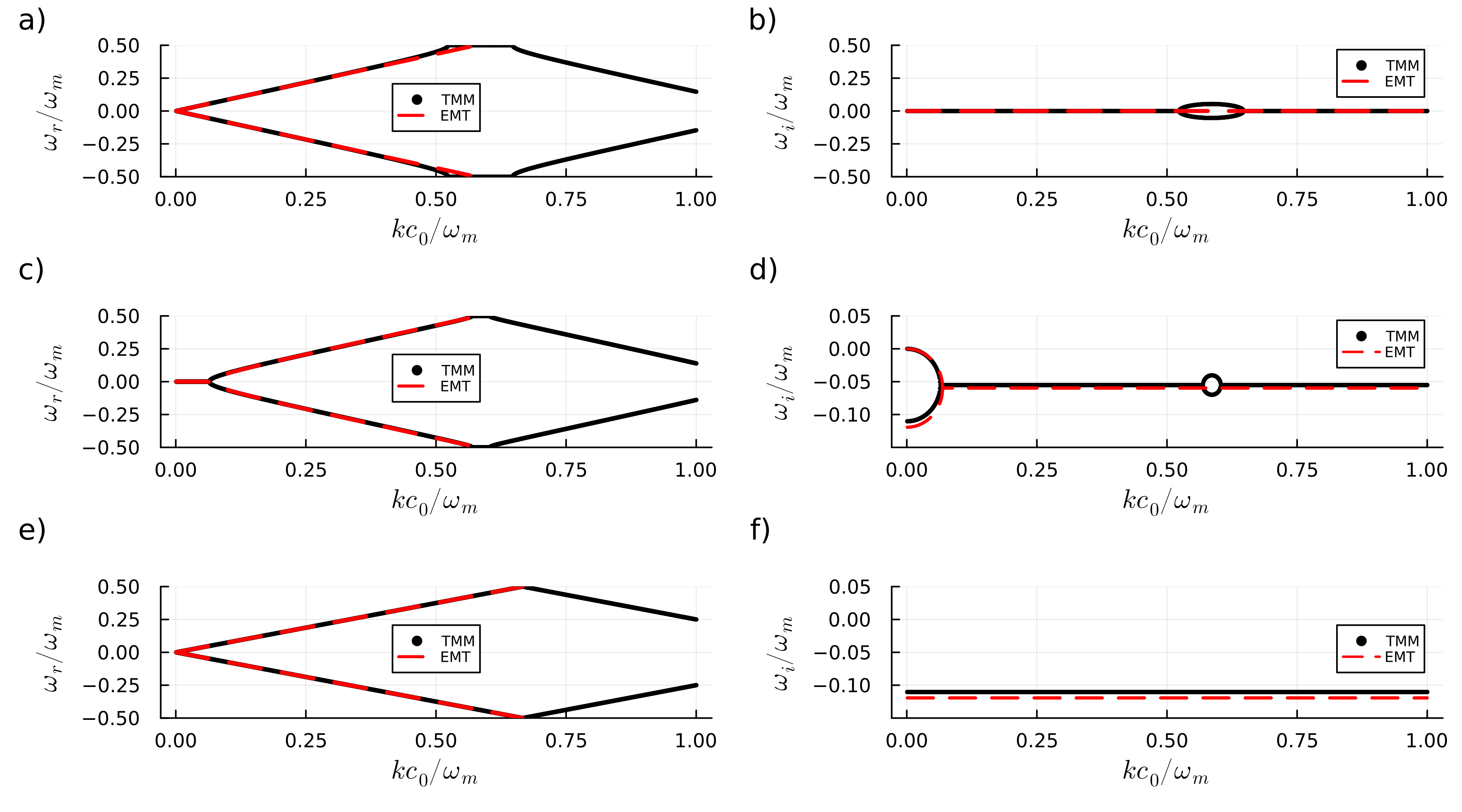}
	\caption{Band diagram of piecewise time-periodic media: a) Band diagram of an active PTC with permittivity modulation; b) temporal decay and growth in the active PTC; c) band diagram of a passive PTC with permittivity modulation; d) temporal decay in the passive PTC; e) band diagram of a passive PTC with impedance-matched modulation; f) temporal decay in the passive PTC with impedance-matched modulation} \label{fig2}
\end{figure*}

The behavior of passive PTCs in comparison with the active PTCs is documented in Fig.~\ref{fig2} with TMM and EMT. The constitutive parameters are assumed to be stepwise-modulated from $1$ to $2$ with a duty cycle of 0.5 and unity whenever time-invariant. An active PTC with permittivity modulation (Figs.~\ref{fig2}a and~\ref{fig2}b) exhibits a small momentum band gap with temporally growing $(\omega_i>0)$ and decaying solutions $(\omega_i<0)$. In the case of passive PTCs with permittivity-only modulation (Figs.~\ref{fig2}c and~\ref{fig2}d), the center Brillouin zone band gap is comparatively smaller; in addition, the dispersion relation exhibits a DC quasi-momentum band gap~\cite{Yin2024}, whose effects can be attributed to the effective electric conductivity introduced by the switching. Per EMT, the dispersion relation of an arbitrary homogenized medium reads as:
\begin{equation}
	\omega^2 \; \varepsilon_{\text{eff}} \; \mu_{\text{eff}} + \omega \; i \left(\frac{\sigma_{\text{eff}}^*\; \varepsilon_\text{eff}}{\mu_0}+\frac{\sigma_\text{eff} \mu_{\text{eff}}}{\varepsilon_0}\right) - \sigma_{\text{eff}} \sigma_{\text{eff}}^* c_0^2 = k^2 c_0^2
\end{equation}
For $\mu_{\text{eff}}=1$ and $\sigma_{\text{eff}}^*=0$, the solutions assume a purely imaginary form for $4k^2 \varepsilon_0\varepsilon_{\text{eff}} \leq \mu_0 \sigma_{\text{eff}}^2$, forming the DC bandgap within this bound while having a stable temporal decay elsewhere; this also reproduces the PTC behavior outside the band gaps. In contrast, the band gaps completely close up in the case of a passive PTC with completely impedance-matched permittivity and permeability modulation, resulting in linear dispersion (Fig.~\ref{fig2}e) and a constant temporal decay (Fig.~\ref{fig2}f). When the temporal interfaces are no longer synchronized, the band gaps other than the DC band gap reemerge due to the reflections from impedance discontinuities within the temporal unit cell.

It should be further stressed that the EMT model invoked herein assumes short and monotonic switching transients compared with the modulation period, and a unit cell that remains electrically small and nondispersive at the highest Floquet harmonic carrying appreciable power. The former assumption concerns the constitutive description of the wave propagation, in which the losses due to finite-duration switching transients~\cite{Maslov2025} are neglected: unlike the cases for permittivity and permeability, the effective conductivity expressions $\sigma_{\text{eff}}$ (Eq.~\ref{eq:emt-3}) and $\sigma_{\text{eff}}^*$ (Eq.~\ref{eq:emt-4}) are insensitive to the shape of the transition. In comparison, the latter assumption is consequential for the physical implementation of photonic temporal interfaces, such as microstrip TL metamaterials periodically loaded with dynamic reactive elements~\cite{Moussa2023,Jones2024}: since these structures are spatially dispersive with a Bragg cutoff determined by the unit cell dimensions, the nondispersive isotropic description is faithful for harmonics residing within this homogenization limit. 

\section{Impedance-matched absorption: results}
In order to test the proposed absorption mechanism and effective medium models, wave interaction with a grounded slab (Dallenbach layer) of thickness $d$ is studied under passive modulation. The half-space bounding the slab is considered to be vacuum, whereas the center frequencies of the normally-incident pulses were assumed to be 200 MHz. The step pulses are approximated with hyperbolic tangents, with the homogenization integrals of Eqs.~\ref{eq:emt-1} and \ref{eq:emt-2} being evaluated for these profiles rather than their step function limit; in addition, a duty cycle of 0.5 as well as a lower permittivity/permeability bound of unity is adopted throughout the analyses, except where noted. The results are verified with semi-analytical Floquet expansion analyses and in-house finite-difference time-domain (FDTD) simulations, whose details are given in Appendices \ref{appendixB} and \ref{appendixC}, respectively.

\subsection{Matching under synchronous modulation}
\begin{figure}[h!]
	\centering 
	\includegraphics[width=.75\textwidth]{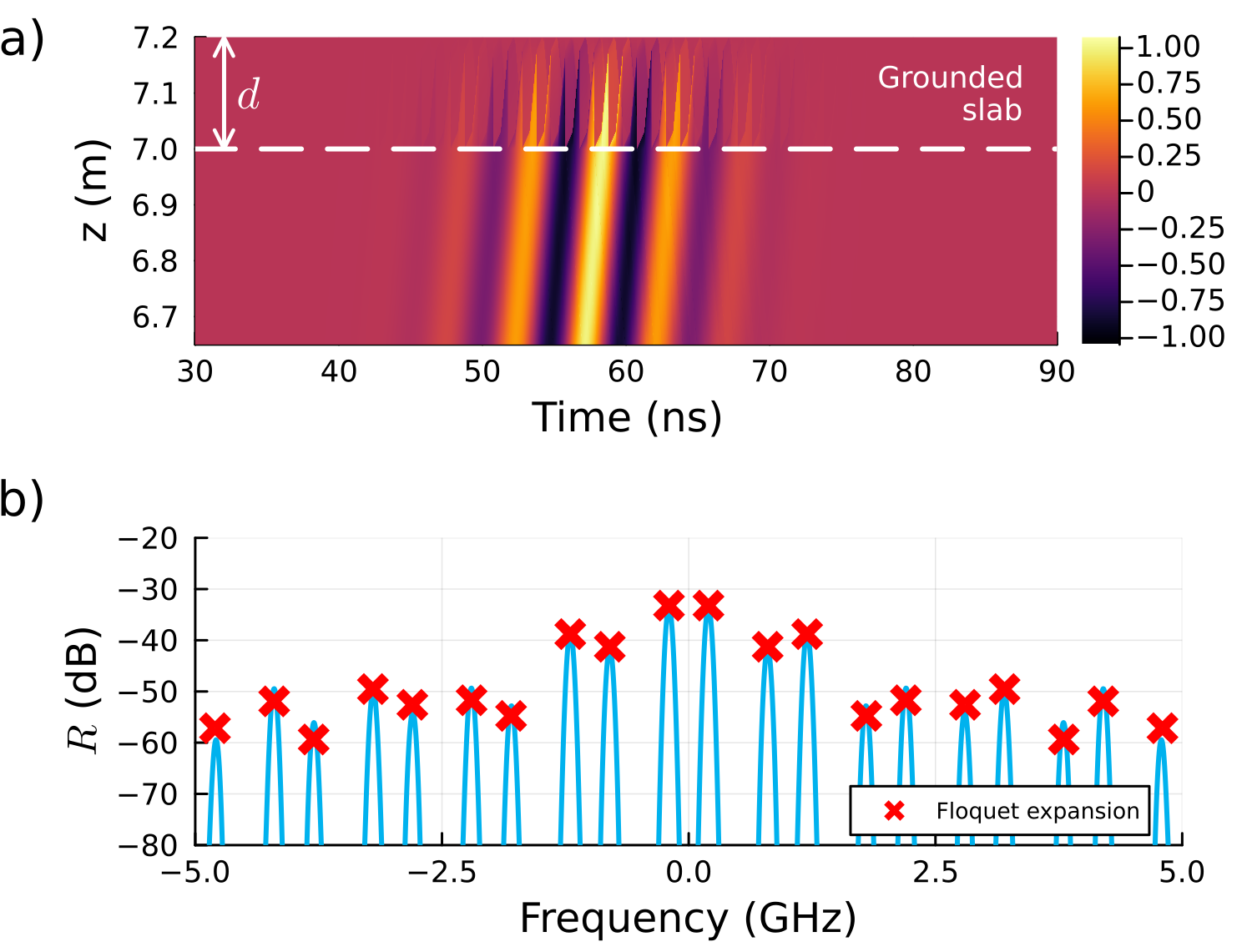}
	\caption{FDTD simulation of reflection from a grounded slab with impedance-matched permittivity and permeability switching: a) Spatiotemporal electric field profile; b) reflectance of Floquet harmonics. } \label{fig3}
\end{figure}

The spacetime profile of the electric field and reflectance of a grounded time-periodic slab of $d=0.2$ m is provided in Fig.~\ref{fig3}. The modulation parameters are chosen as $\Delta \varepsilon=\Delta \mu=4$ with $T_e=T_m=1$ ns. Both modulation waveforms are synchronous $(\varepsilon(t)=\mu(t))$, which denotes a purely impedance-matched temporal metamaterial. The results for an incident Gaussian pulse demonstrate efficient absorption of the waves: in Fig.~\ref{fig3}a, it can be observed that the incident wave encounters no significant impedance mismatch at the slab boundary at $z=0$ and electric fields within the structure are rapidly attenuated at the falling edges of the modulation. Furthermore, reflectance of the Floquet harmonics is in line with the semi-analytical calculations, showcasing negligible higher-order reflections. 

\subsection{Absorption performance and the Rozanov bound}
The performance of a grounded slab is governed by a fundamental trade-off between absorption efficiency, bandwidth, and thickness. This theoretical limit is quantified by the Rozanov bound~\cite{Rozanov2000}, which defines the maximum achievable absorption for a given thickness and bandwidth:
\begin{align}
	\left|\int_0^\infty d\lambda \ln |\Gamma(\lambda)| \right| \leq 2\pi^2 \mu_r d \label{eq:rozanov}
\end{align}
where $|\Gamma(\lambda)|$ is the wavelength-dependent reflection magnitude of the structure, $\mu_r$ is the \textit{static} relative permeability and $d$ is the thickness. In practice, this bound, as well as the closely-related Bode-Fano limit, can be bypassed by linear time-varying systems using single-shot switching~\cite{Shlivinski2018,Li2020,Firestein2022,Yang2022} or periodic modulations~\cite{Chambers2005,Mostafa2022,Hayran2024,Ciabattoni2025}. Thus, it is imperative to benchmark the absorption performance of the proposed mechanism in comparison with this fundamental limit.

In periodically time-varying systems, energy can be scattered into Floquet harmonics far outside the incident frequency band. Consequently, evaluating absorption efficiency based on a fully-defined incident spectrum is more physically insightful than relying on the traditional Rozanov integral (Eq.~\ref{eq:rozanov}). This figure-of-merit is expressed as~\cite{Firestein2022,Hayran2024,Ciabattoni2025}:
\begin{align}
	\eta_{\text{abs}}&=1-\frac{\int_{-\infty}^{\infty} d\omega |E_{\text{ref}}(\omega)|^2 }{\int_{-\infty}^{\infty} d\omega |E_{\text{inc}}(\omega)|^2} \label{eq:absorption_efficiency}
\end{align}
where $E_{\text{ref}}$ and $E_{\text{inc}}$ are the reflected and incident electric field spectra, respectively. Since the original Rozanov bound is defined for band-limited signals in the wavelength space, rather than the frequency space as in Eq.~\ref{eq:absorption_efficiency}, theoretical absorption limit is obtained numerically from the variational approach reported by Ciabattoni \emph{et al.}~\cite{Ciabattoni2025}. As the permeability of analyzed structure fluctuates in time, this absorption bound is evaluated using its peak value to establish a stringent and conservative upper limit. The incident field is configured to have a flat and band-limited spectrum within the frequency space with $f_c=200$ MHz and $\Delta f=300$ MHz.

\begin{figure}[h!]
	\centering 
	\includegraphics[width=0.85\textwidth]{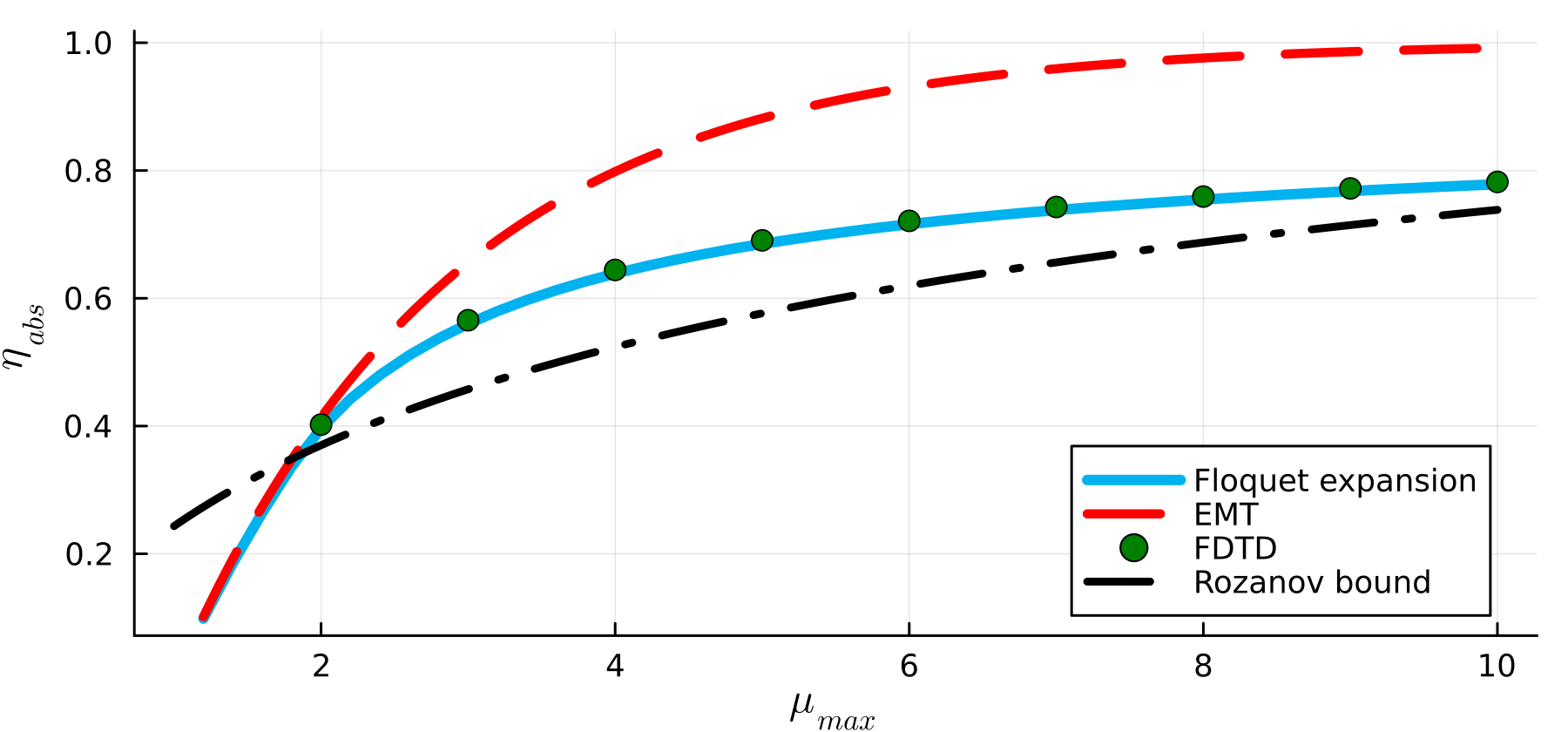}
	\caption{Absorption efficiency of a band limited signal with respect to the maximum permeability during modulation for an  electrically thin grounded slab of $d=0.01$ m.} \label{fig6}
\end{figure}

Fig.~\ref{fig6} illustrates the semi-analytically and numerically obtained absorption efficiencies alongside the Rozanov bound as well as the idealized results under EMT for an electrically thin slab of $0.01$ m length. The switching period for this example is chosen to be $T_e=T_m=0.25$ ns to accentuate the modulation-induced effective conductivities within the layer; such sub-nanosecond switching speeds are physically realizable using high-speed step recovery diodes~\cite{Yang2022}. Noticeably, the proposed mechanism exceeds the Rozanov bound for certain modulation depths. The static EMT results, representing an idealized PML, also surpass this theoretical bound significantly; this is physically permissible since effective magnetic losses, being absent from Eq.~\ref{eq:rozanov}, remove the traditional upper limit on maximal absorption~\cite{Firestein2023}. As the modulation depth increases, the calculated efficiency begins to deviate from the idealized EMT predictions, approaching the Rozanov bound. This can be attributed to the fact that homogenization overestimates the loss for larger $\Delta\varepsilon$ and $\Delta\mu$. The absorption efficiency can further be increased under shorter switching periods, enabling beyond Rozanov performance with much smaller modulation depths.  

\subsection{Matching under asynchronous modulation}
\begin{figure}[h!]
	\centering 
	\includegraphics[width=.95\textwidth]{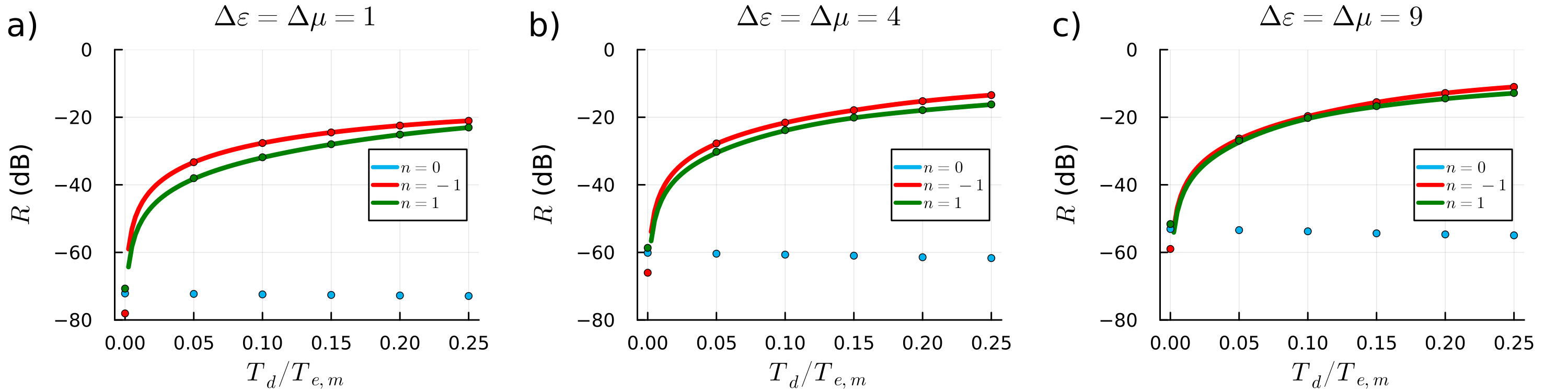}
	\caption{Reflectance of a time-varying half-space with respect to the time delay $(T_d)$ between permittivity and permeability modulations and the modulation depth. Line and scatter plots show the Floquet expansion and FDTD results, respectively.} \label{fig4}
\end{figure}

In addition to synchronous impedance-matched modulation, the absorption performance under asynchronicity is also evaluated. The initial results are obtained for a half-space of switched medium to test the bounds of matching at the interface; asynchronicity for this specific case is achieved through introducing a time delay $(T_d)$ between the permittivity and permeability functions. Reflectance data for the lowest-order harmonics is given in Fig.~\ref{fig4} for $T_e=T_m=1$ ns. The FDTD results are obtained for a sufficiently thick slab of 15 m, ensuring convergence to the half-space results. Floquet expansion calculations predict essentially no reflection for the zeroth-order harmonic, whereas FDTD results exhibit a relatively small but finite reflection: this can be attributed to leapfrog offset errors introduced by the Yee cell and decrease for smaller meshes. Evidently, interfacial reflection for lowest-order nonzero harmonics increases for longer time-delays. This behavior originates from instantaneous impedance discontinuities occurring within the temporal unit cell~\cite{GaxiolaLuna2021}. For a matched medium under synchronous modulation, the effective impedances seen by the Floquet modes are equivalent to the instantaneous wave impedance, which is unaltered at all times. When this staticity is broken by a time-delay, the higher-order modes encounter different characteristic impedances within the layer, which results in reflections. It is further observed that this mismatch is exacerbated by stronger modulation profiles, necessitating the use of alternative approaches in addition to the simple matching condition~(Eq.~\ref{eq:emt_match}).

\begin{figure}[h!]
	\centering 
	\includegraphics[width=0.95\textwidth]{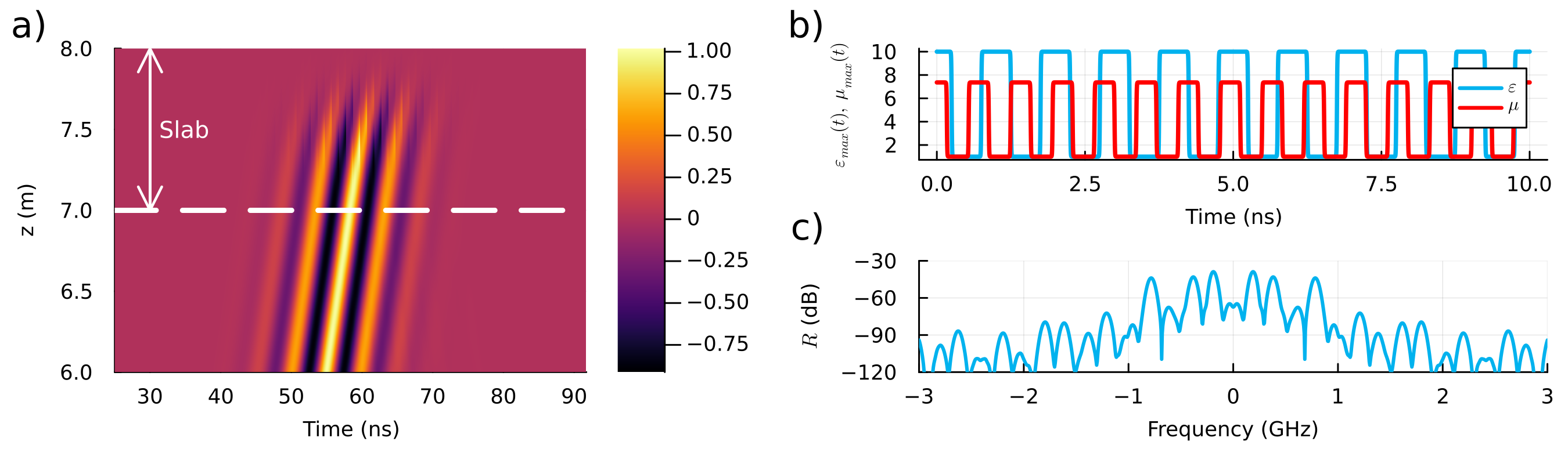}
	\caption{FDTD simulation of a grounded and tapered slab with asynchronous modulation and an almost-periodic instantaneous wave impedance: a) Spatiotemporal electric field profile; b) Time-domain permittivity and permeability functions at $z=8$ m; c) spectral reflectance.} \label{fig5}
\end{figure}

Nevertheless, impedance matching under the approximate relation in Eq.~\ref{eq:emt_match} can still be ensured as long as excitation of higher-order harmonics at the interface is sufficiently suppressed. This can be done through tapering the modulation depth across the layer, which itself is a widely-adopted technique for the numerical implementation of PMLs in the FDTD method~\cite{Berenger1994}. In order to test the bounds of tapered matching, the periods of permittivity and permeability functions are fixed to $1$ ns and $1/\sqrt{2}$ ns, respectively; this results in an almost-periodic PTC~\cite{Koufidis2024} with a strong detuning between the permittivity and permeability profile. The modulation depth of the permittivity is graded from zero to a maximum value of 9 at the end of a slab of 1 m with a cubic polynomial. The corresponding duty cycle and modulation depth of the permeability function is obtained from the quasistatic matching condition in Eq.~\ref{eq:emt_match}. FDTD results are given in Fig.~\ref{fig5}; these suggest that absorption levels similar to that of the synchronous switching can be obtained at the expense of a longer taper and a larger dielectric modulation depth. Furthermore, this also shows that bounds on the permeability change can be relaxed under a higher switching frequency. 

\subsection{Reconfigurable material responses in the effective medium limit}
Besides the matching mechanism discussed so far, passive temporal metamaterials under permittivity or permeability switching can be used to engineer arbitrary lossy material responses at desired frequencies. Validation of this effect is provided by the same grounded slab configuration as before, assuming purely dielectric material properties. Using TL theory~\cite{Kinayman2005}, matching condition of this structure in the effective medium limit is given as:
\begin{align}
	\sqrt{\varepsilon_c} \cot\left( \frac{\omega}{c_0}\sqrt{\varepsilon_c} d\right)+i=0
\end{align} 
where $\varepsilon_c$ and $d$ are the complex relative permittivity and thickness, respectively. For $d=0.2$ m, the target complex permittivity is obtained as $\varepsilon_c=3.913+i 2.337$ through numerical optimization~\cite{Mogensen2018}. Per Eq.~\ref{eq:emt-3}, the effective conductivity $\sigma_{\text{eff}}$ and the necessary modulation depth $\Delta \varepsilon$ are given as $\text{Im}\left[\varepsilon_c\right]\omega\varepsilon_0$ and $\text{Im}\left[\varepsilon_c\right]\omega T_e$, respectively. Using the effective dielectric constant expression (Eq.~\ref{eq:emt-1}), the real part of the complex permittivity $(\text{Re}\left[\varepsilon_c\right])$ can be adjusted through calibrating the bounds of the modulation and the duty cycle.
\begin{figure}[h!]
	\centering 
	\includegraphics[width=0.95\textwidth]{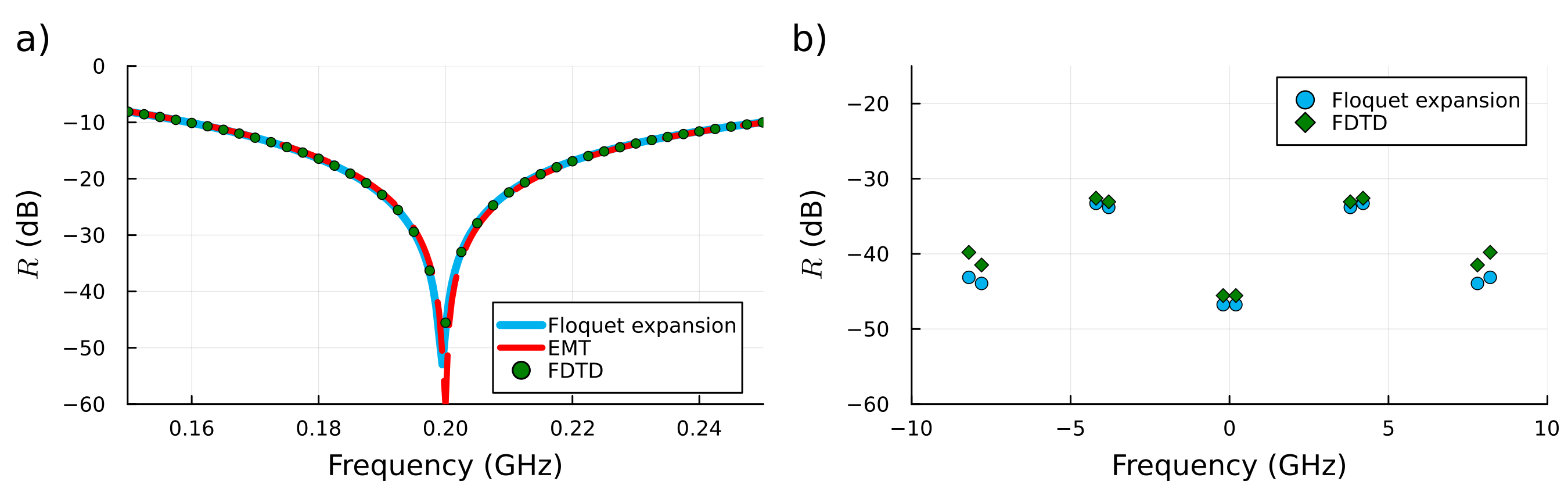}
	\caption{Reflectance of a grounded slab with passive permittivity modulation: a) Zeroth-order reflectance; b) reflectance of Floquet harmonics for $\omega_\text{inc}=2\pi \cdot 0.2$ rad/ns.} \label{fig7}
\end{figure}

Reflectance of the matched slab within the effective medium limit is given in Fig.~\ref{fig7}; the modulation parameters were fixed to $T_e=0.25$ ns, $\varepsilon_1=3.546$, $\varepsilon_2=4.28$ and $\tau_{1e}=0.45 T_e$. The effective static response shows excellent agreement with both Floquet expansion and FDTD results, validating the use of passive switching for reconfigurable lossy materials. Reflectance of the first-order harmonics remains relatively small despite being higher than that of the zeroth-order harmonic. This discrepancy is largely governed by the relatively large modulation depth, which can be further relaxed by increasing the modulation frequency.

\section{Limitations on temporal uniaxial perfectly matched layers}
Since the impedance-matched medium explored herein so far can only absorb TEM waves, a question arises on the applicability of the matching condition to a realistic slab configuration where the incident field can come from an arbitrary angle and in different polarizations. This problem has historically driven the development of a "true" PML medium for computational electromagnetics~\cite{Berenger1994,Sacks1995,Gedney1996}, as well as the quest for its realistic physical implementation~\cite{Ziolkowski1997,Teixeira2003,Valagiannopoulos2015}. Focusing solely on TE-polarized plane waves, the matching condition along the $z$-axis can be summarized as~\cite{Sacks1995}:
\begin{align}
	\varepsilon_{xx}=\mu_{yy}=\frac{1}{\mu_{zz}}
\end{align}
where the constitutive parameters $\varepsilon_{xx}$ and $\mu_{yy}$ are lossy. Evidently, this requires $\mu_{zz}$ to be active; it exhibits a magnetic Debye response with a negative coupling term and is expected to be unstable without the presence of loss in the transverse permittivity and permeabilities~\cite{Ziolkowski1997}.

Owing to the linear time-variant nature of the materials considered herein, the usual matching condition needs to be reevaluated, which can be done under the operator description of the constitutive parameters under a Fourier basis~\cite{Horsley2023,Hooper2025,Iplikcioglu2025}. Considering the problem of wave reflection from an interface between vacuum and an arbitrarily time-varying uniaxial medium as in the uniaxial PML approach~\cite{Sacks1995,Gedney1996}, we can write the dispersion relation of TE waves in the latter as (Appendix \ref{appendixD}):
\begin{align}
	\left[\hat{k}_0 \hat{\mu}_{zz} \hat{k}_0 \right]^{-1} \hat{k}_y^2 + \left[\hat{k}_0 \hat{\mu}_{yy} \hat{k}_0 \right]^{-1} \hat{k}_{zp}^2 = \hat{\varepsilon}_{xx} \label{eq:uniaxial-dispersion}
\end{align}
where the hat symbol denotes the operator form. $\hat{k}_y$ corresponds to the transverse wavevector, which is shared across the both half-spaces. In turn, $\hat{k}_{zp}$ denotes the $z$-oriented wavevector in the uniaxial half-space. Per the reflection operator of time-varying half-spaces (Appendix \ref{appendixB}), complete admittance matching is achieved under the condition: 
\begin{align}
	\hat{Y}_0 = \hat{Y}_{p}
\end{align} 
where $\hat{Y}_0=\left[\hat{\omega} \mu_0\right]^{-1} \hat{k}_{z i}$ and $\hat{Y}_{p}=\left[\hat{\omega }\mu_0 \hat{\mu}_{yy}\right]^{-1} \hat{k}_{z p}$ are the admittance operators for multifrequency waves, $\hat{k}_0=\hat{\omega}/c_0$ and $\hat{k}_{z i}=\left[\hat{k}_0^2-\hat{k}_y^2\right]^{1/2}$; note that $\hat{Y}_0$ and $\hat{k}_y$ are multiplication operators and correspond to diagonal matrices in matrix algebra, since the incident wave impinges from a linear time-invariant half-space. The exact matching requires:
\begin{align}
	\hat{k}_{z p} = \hat{k}_0 \hat{\mu}_{yy} \hat{k}_0^{-1} \hat{k}_{z i} \label{eq:operator-match}
\end{align}
Inserting Eq.~\ref{eq:operator-match} into Eq.~\ref{eq:uniaxial-dispersion} and noting that $\hat{k}_y^2=\hat{k}_0^2-\hat{k}_{z i}^2$, we obtain:
\begin{align}
	\hat{k}_0^{-1} \hat{\mu}_{zz}^{-1} \hat{k}_0 -\hat{k}_0^{-1} \hat{\mu}_{zz}^{-1} \hat{k}_0^{-1} \hat{k}_{z i}^2 + \hat{k}_0^{-2} \hat{k}_{z i} \hat{k}_0 \hat{\mu}_{yy} \hat{k}_0^{-1} \hat{k}_{z i} = \hat{\varepsilon}_{xx}
\end{align}
It can be seen that the usual matching condition $\hat{\varepsilon}_{xx}=\hat{\mu}_{yy}=\hat{\mu}_{zz}^{-1}$ is guaranteed to exactly satisfy the dispersion equation for an arbitrary $\hat{k}_{z i}$ only when the waves are normally incident $(\hat{k}_{z i}=\hat{k}_0)$ or when the constitutive parameters are commuting multipliers, i.e. under a linear time-invariant material response. While the non-commutativity of these operators under arbitrary time-modulation does not preclude the existence of matched absorption at arbitrary angles, it implies that such conditions are fundamentally frequency- and modulation-dependent, potentially leading to inexact matching for Floquet harmonics. Thus, two-dimensional impedance matching may necessitate the use of active components alongside inverse-designed modulation profiles~\cite{Garg2025}.

\section{Conclusion}
In conclusion, it is demonstrated that a temporal metamaterial can be engineered to behave as a PML-like matched absorber under passive modulation of the constitutive parameters. An effective medium model identifies modulation-dependent dielectric and magnetic losses in the quasistatic limit, which can be tuned to achieve reconfigurable operation as well as operation under asynchronous modulation. The proposed mechanism is also shown to surpass the Rozanov bound, with absorption efficiencies exceeding this limit being obtained for an electrically-thin grounded slab. Separately, an operator analysis of the generalized two-dimensional problem for waves with arbitrary transverse wavevectors reveals the matching condition to be intrinsically modulation- and frequency-dependent for linear time-variant media. 

Beyond the planar microstrip configuration, these switched metamaterials can be constructed in the form of stacked metasurfaces~\cite{Movahediqomi2026}, using programmable capacitors and high-frequency switches to obtain complex material responses. With the recent advances in the temporal permeability modulation of ferromagnetic materials at microwave frequencies~\cite{Kodama2023}, the absorption scheme can possibly be implemented in bulk metamaterials as well. While optical implementation of this PML-like matched absorption is complicated by the lack of intrinsic magnetic material responses in these frequencies, the proposed mechanism can possibly be approximated using graphene-based magnetic metamaterials~\cite{Papasimakis2013} under passive carrier density modulation~\cite{Maslov2018}.

\section*{Acknowledgments}
The author thanks Prof. M. I. Aksun for useful discussions.

\appendix
\section*{Appendix}
\section{Transfer matrix method}\label{appendixA}
Temporally-discrete and spatially-infinite media are most effectively analyzed using the transfer-matrix method (TMM)~\cite{Asgari2024}. For the sake of brevity, only the derivation for the passive and synchronized PTCs is documented. Modulation functions for a two-phase and homogeneous time-periodic medium can be written as:
\begin{align}
	\varepsilon(t) &= \begin{cases}
		\varepsilon_1, & n(\tau_1+\tau_2)<t< n(\tau_1+\tau_2) + \tau_1,  \\
		\varepsilon_2, & n(\tau_1+\tau_2)-\tau_2 <t< n(\tau_1+\tau_2),
	\end{cases} \\
	\mu(t) &= \begin{cases}
		\mu_1, & n(\tau_1+\tau_2)<t< n(\tau_1+\tau_2) + \tau_1,  \\
		\mu_2, & n(\tau_1+\tau_2)-\tau_2 <t< n(\tau_1+\tau_2),
	\end{cases}
\end{align}
for $n \in \mathbb{Z}$. Under the assumption of $\varepsilon_2 > \varepsilon_1$ and $\mu_2 > \mu_1$, the following definitions of the transfer matrix elements are employed for a fixed wavenumber $k$~\cite{Ramaccia2021}:
\begin{align}
	M_{21} = & 
	\begin{bmatrix}
		t^{D,B}_{21} & r^{D,B}_{21} \\
		r^{D,B}_{21} & t^{D,B}_{21}
	\end{bmatrix} \\
	M_{12} = & 
	\begin{bmatrix}
		t^{E,H}_{12} & r^{E,H}_{12} \\
		r^{E,H}_{12} & t^{E,H}_{12}
	\end{bmatrix} \\
	P_1 = & 
	\begin{bmatrix}
		e^{-i\frac{k c_0}{\sqrt{\varepsilon_1 \mu_1}} \tau_1} & 0 \\
		0 & e^{i\frac{k c_0}{\sqrt{\varepsilon_1 \mu_1}} \tau_1}
	\end{bmatrix} \\
	P_2 = & 
	\begin{bmatrix}
		e^{-i\frac{k c_0}{\sqrt{\varepsilon_2 \mu_2}} \tau_2} & 0 \\
		0 & e^{i\frac{k c_0}{\sqrt{\varepsilon_2 \mu_2}} \tau_2}
	\end{bmatrix}
\end{align}
where the reflection and transmission coefficients $r$ and $t$ are found either for flux continuity for rising edges $(D,B)$ or field continuity for falling edges $(E,H)$~\cite{Xiao2014,Morgenthaler1958}:
\begin{align}
	r_{21}^{D} &= \frac{1}{2} \left[\frac{\varepsilon_1}{\varepsilon_2} - \sqrt{\frac{\varepsilon_1 \mu_1}{\varepsilon_2 \mu_2}} \right] \\
	t_{21}^{D} &= \frac{1}{2} \left[\frac{\varepsilon_1}{\varepsilon_2} + \sqrt{\frac{\varepsilon_1 \mu_1}{\varepsilon_2 \mu_2}} \right] \\
	r_{12}^{E} &= \frac{1}{2} \left[1 - \sqrt{\frac{\mu_1 \varepsilon_2}{\mu_2 \varepsilon_1}} \right] \\
	t_{12}^{E} &= \frac{1}{2} \left[1 + \sqrt{\frac{\mu_1 \varepsilon_2}{\mu_2 \varepsilon_1}} \right]
\end{align}
Having defined the transfer functions, we can obtain the band diagram from the following eigenproblem:
\begin{align}
	P_1 M_{21} P_2 M_{12} 
	\begin{bmatrix}
		E^+ \\
		E^-
	\end{bmatrix} = e^{-i \omega (\tau_1 + \tau_2)}
	\begin{bmatrix}
		E^+ \\
		E^-
	\end{bmatrix} \label{eq:eigenproblem}
\end{align}
For the specific case of impedance-matched modulation with $\varepsilon_1=\mu_1$ and  $\varepsilon_2=\mu_2$, the eigenproblem can be written as:
\begin{align}
	P_1 M_{21} P_2 M_{12}
	\begin{bmatrix}
		E^+ \\
		E^-
	\end{bmatrix} =
	\begin{bmatrix}
		\frac{\varepsilon_1}{\varepsilon_2} e^{-ik c_0 \left(\frac{\tau_1}{\varepsilon_1} +\frac{\tau_2}{\varepsilon_2} \right)} & 0 \\
		0 & \frac{\varepsilon_1}{\varepsilon_2} e^{ik c_0 \left(\frac{\tau_1}{\varepsilon_1} +\frac{\tau_2}{\varepsilon_2} \right)} \\
	\end{bmatrix} 
	\begin{bmatrix}
		E^+ \\
		E^-
	\end{bmatrix} = e^{-i \omega (\tau_1 + \tau_2)}
	\begin{bmatrix}
		E^+ \\
		E^-
	\end{bmatrix} 
\end{align}
Thus, the dispersion relation can be reduced to Eq.~\ref{eq:ptc_matched_dispersion}.

\section{Floquet expansion method}\label{appendixB}
Since the temporal modulation considered herein is purely periodic, the reflection of electromagnetic waves from a time-variant isotropic slab of length $d$ can be calculated semi-analytically through Floquet expansion~\cite{Asgari2024} within an operator formalism~\cite{Horsley2023,Hooper2025,Iplikcioglu2025}. Under this approach, standard scalar wave parameters can be replaced by equivalent operator expressions that act on the spectrum of the electromagnetic fields. This enables the robust time-harmonic representation of multi-frequency coupling in a matrix form within the Floquet basis, resulting in expressions similar to those by Ciabattoni \emph{et al.}~\cite{Ciabattoni2025}. As an example, double-time convolution involving a dynamic susceptibility function can be represented in the operator form as:
\begin{align}
	\textbf{P}(t)=\varepsilon_0 \int_{-\infty}^{\infty} dt' \chi(t,t-t') \textbf{E}(t') \rightarrow \textbf{P} = \varepsilon_0 \hat{\chi} \textbf{E}
\end{align}
where the hat symbol denotes the operator (or matrix) form. It is worth noting that an antinormal-like operator ordering has been assumed~\cite{Horsley2023}.

It is assumed that a time-variant slab of thickness $d$ is grounded at $z=d$ with a perfect electrical conductor; a $x$-polarized incident plane wave that is traveling in $z$-direction impinges on the structure. Consequently, the transverse fields in the space below $(1)$ and within the slab $(2)$ can be written as~\cite{Horsley2023}:
\begin{align}
	E_x^{(1)} = e^{i \hat{k}_{z1} z} + e^{-i \hat{k}_{z1} z} \; \hat{\Gamma}, &\;\;\;\;\;\; (z \leq 0)\\
	E_x^{(2)} = e^{i \hat{k}_{z2} z} \hat{A} + e^{-i \hat{k}_{z2} z} \hat{B},   & \;\;\;\;\;\; (z>0) \\
	H_y^{(1)} = \hat{Y}_1 \left[e^{i \hat{k}_{z1} z} - e^{-i \hat{k}_{z1} z} \hat{\Gamma}\right],  &\;\;\;\;\;\; (z \leq 0)\\
	H_y^{(2)} = \hat{Y}_2 \left[e^{i \hat{k}_{z2} z} \hat{A} - e^{-i \hat{k}_{z2} z}  \hat{B}\right],  &\;\;\;\;\;\; (z>0)  
\end{align}
where $\hat{Y}_n = \left[\hat{\omega}\mu_0 \hat{\mu}_c\right]^{-1} \hat{k}_{zn}$ is the admittance operator and $\hat{k}_{zn}=\left[\hat{\omega} \hat{\mu}_c \hat{\omega} \hat{\varepsilon}_c \left(\mu_0 \varepsilon_0\right) - k_y^2 I\right]^{1/2}$ is the longitudinal wavevector operator for a fixed $k_y$ vector. $\hat{\varepsilon}_c$ and $ \hat{\mu}_c$ are the complex permittivity and permeability operators:
\begin{align}
	\hat{\varepsilon}_c &= \hat{\varepsilon} + i \hat{\omega}^{-1} \varepsilon_0^{-1} \hat{\sigma}\\
	\hat{\mu}_c &= \hat{\mu} + i \hat{\omega}^{-1} \mu_0^{-1} \hat{\sigma}^*
\end{align}
with $\hat{\sigma}$ and $\hat{\sigma}^*$ being electric and magnetic conductivity operators. These are practically constructed in the form of Toeplitz matrices using fast Fourier transform of the time-domain waveforms, with the conductivity profiles being obtained from Eqs.~\ref{eq:smooth-1e} to ~\ref{eq:smooth-2m}. At $z=d$ perfect conductor boundary conditions are applied:
\begin{align}
	E_x^{(2)} &= e^{i \hat{k}_{z2} d} \hat{A} + e^{-i \hat{k}_{z2} d} \hat{B}=0\\
	\hat{B} &=-e^{i \hat{k}_{z2} 2 d} \hat{A}
\end{align}
Enforcing the field continuity at $z=0$,
\begin{align}
	I + \hat{\Gamma}=\hat{A} + \hat{B}&=\left[I-e^{i \hat{k}_{z2} 2 d}\right]\hat{A}\\
	\hat{Y}_1 \left[I - \hat{\Gamma}\right] &= \hat{Y}_2 \left[\hat{A} - \hat{B}\right]=\hat{Y}_2 \left[I + e^{i \hat{k}_{z2} 2 d} \right] \hat{A}
\end{align}
Algebraic manipulation of the expressions above yields the reflection operator for the grounded slab:
\begin{align}
	\hat{\Gamma} = \left[\hat{Y}_1 + \hat{Y}_2 \hat{X}\right]^{-1} \left[\hat{Y}_1 - \hat{Y}_2 \hat{X}\right]
\end{align}
where $\hat{X}=\left[I + e^{i \hat{k}_{z2} 2 d} \right] \left[I-e^{i \hat{k}_{z2} 2 d}\right]^{-1}$. The reflection coefficient operator for the time-modulated half-space $(\hat{\Gamma}_0)$ can be retrieved for the limit of an infinitely-thick slab: 
\begin{align}
	\hat{\Gamma}_0 = \lim_{d \rightarrow \infty} \hat{\Gamma} = \left[\hat{Y}_1 + \hat{Y}_2\right]^{-1} \left[\hat{Y}_1 - \hat{Y}_2 \right]
\end{align}
It is worth highlighting that for this specific case, the matrix square root should be computed in a way that enforces radiation condition at infinity~\cite{Horsley2023}, which is handled here via the Schur method~\cite{Bjorck1983}. While the calculations herein were performed for isotropic materials, the expressions themselves are valid for uniaxial slabs that support TE modes, as long as the corresponding wavevector and admittance operators are derived using the appropriate dispersion relation~(Appendix~\ref{appendixD}).

\section{FDTD simulations}\label{appendixC}
Update equations for the 1D FDTD simulations are given as~\cite{Taflove2005,Liu2007}:
\begin{align}
	E_x^{n+1}(i) =& \left[\frac{\varepsilon^n(i)-\frac{\sigma_\text{PML} (i) \Delta t}{2\varepsilon_0}}{\varepsilon^{n+1}(i)+\frac{\sigma_\text{PML} (i) \Delta t}{2\varepsilon_0}}\right] E_x^{n}(i) - \frac{\Delta t}{\Delta z \;  \varepsilon_0 }\left[\frac{1}{\varepsilon^{n+1}(i)+\frac{\sigma_\text{PML} (i) \Delta t}{2\varepsilon_0}}\right] \times \nonumber \\
	& \left[H_y^{n+0.5}\left(i+\frac{1}{2}\right)-H_y^{n+0.5}\left(i-\frac{1}{2}\right)\right]\\
	H_y^{n+0.5}\left(i+\frac{1}{2}\right)=&\left[\frac{\mu^{n-0.5}(i+0.5)-\frac{\sigma_\text{PML}^* (i+0.5) \Delta t}{2\mu_0}}{\mu^{n+0.5}(i+0.5)+\frac{\sigma_\text{PML}^* (i+0.5) \Delta t}{2\mu_0}}\right] \times 
	H_y^{n-0.5}\left(i+\frac{1}{2}\right) \nonumber \\
	&-\frac{\Delta t}{\Delta z \; \mu_0} \times  \left[\frac{1}{\mu^{n+0.5}(i+0.5)+\frac{\sigma_\text{PML}^* (i+0.5) \Delta t}{2\mu_0}}\right] \times  \left[E_x^{n}(i+1)-E_x^{n}(i)\right]
\end{align}
where $\sigma_\text{PML}$ and $\sigma_\text{PML}^*$ are the tapered electric and magnetic conductivities to enforce PMLs. $\Delta t$ and $\Delta z$ correspond to temporal and spatial step sizes, respectively. The simulation region is terminated at both ends with perfect conductor boundaries. Since standard update equations for nondispersive temporal modulation naturally enforce the continuity of $\textbf{D}$ and $\textbf{B}$~\cite{Bahrami2024}, the following corrections are applied on the field components at the falling edges of time-dependent permittivity and permeability functions ($\varepsilon^{n+1}<\varepsilon^{n}$ or $\mu^{n+0.5}<\mu^{n-0.5}$):
\begin{align}
	\tilde{E}_x^{n+1}(i)&=\frac{\varepsilon^{n+1}(i)}{\varepsilon^n(i)} {E}_x^{n+1}(i)\\
	\tilde{H}_y^{n+0.5}\left(i+\frac{1}{2}\right)&=\frac{\mu^{n+0.5}\left(i+\frac{1}{2}\right)}{\mu^{n-0.5}\left(i+\frac{1}{2}\right)} {H}_y^{n+0.5}\left(i+\frac{1}{2}\right)
\end{align}
The CFL factor and the spatial step size are taken to be 0.5 and 250 $\mu$m, respectively.
\section{Waves in time-varying uniaxial media}\label{appendixD}
Under the operator formalism~(Appendix \ref{appendixB}), Maxwell's equations under TE polarization for an arbitrarily time-varying uniaxial medium can be written as:
\begin{align}
	\frac{d E_x}{dy} & = -i \hat{\omega}\mu_0 \hat{\mu}_{zz} H_z \label{eq:maxwell1} \\
	\frac{d E_x}{dz} & = i \hat{\omega}\mu_0 \hat{\mu}_{yy} H_y \label{eq:maxwell2} \\
	\frac{d H_z}{dy} - \frac{d H_y}{dz} & = -i\hat{\omega} \varepsilon_0 \hat{\varepsilon}_{xx} E_x \label{eq:maxwell3}
\end{align}
where $-i\hat{\omega}$, $\hat{\varepsilon}$ and $\hat{\mu}$ are the derivative, permittivity and permeability operators, respectively. The medium is assumed to be translationally-invariant along the $x$-axis. Substituting Eqs.~\ref{eq:maxwell1} and \ref{eq:maxwell2} into Eq.~\ref{eq:maxwell3}, and seeking out plane wave solutions $(E_x \propto e^{i \hat{k}_y y + i \hat{k}_z z})$, we obtain:
\begin{align}
	\frac{d}{dy} \left(\left[-i \hat{\omega}\mu_0 \hat{\mu}_{zz}\right]^{-1} \frac{d E_x}{dy}\right) - 
	\frac{d}{dz} \left( \left[i\hat{\omega}\mu_0 \hat{\mu}_{yy}\right]^{-1} \frac{d E_x}{dz}\right) & = -i\hat{\omega} \varepsilon_0 \hat{\varepsilon}_{xx} E_x \\
	\left(\left[\hat{\omega}\mu_0 \hat{\mu}_{zz}\right]^{-1} \hat{k}_y^2 +  \left[\hat{\omega}\mu_0 \hat{\mu}_{yy}\right]^{-1} \hat{k}_z^2 \right) E_x & = \hat{\omega} \varepsilon_0 \hat{\varepsilon}_{xx} E_x
\end{align}
with $\hat{k}_{y,z}$ being the wavevector operators. Dividing both sides by $\hat{\omega}\varepsilon_0$ and defining the wavenumber operator as $\hat{k}_0=\hat{\omega}\sqrt{\varepsilon_0 \mu_0}$, the dispersion relation of plane waves in uniaxial time-varying media is derived:
\begin{align}
	\left[\hat{k}_0 \hat{\mu}_{zz} \hat{k}_0\right]^{-1} \hat{k}_y^2 + \left[\hat{k}_0 \hat{\mu}_{yy} \hat{k}_0\right]^{-1} \hat{k}_z^2 = \hat{\varepsilon}_{xx} \label{eq:uniaxial-dispersion2}
\end{align}

\bibliographystyle{unsrt}
\bibliography{references}
\end{document}